\documentclass[]{aastex631}

\begin{document}

\title{The GMRT High Resolution Southern Sky Survey for pulsars and transients -- IV: Discovery of 4 new pulsars with an FFA search}

\author{S.~Singh}
\affiliation{National Centre for Radio Astrophysics, Tata
Institute of Fundamental Research, Pune 411 007, India}

\author{J.~Roy}
\affiliation{National Centre for Radio Astrophysics, Tata
Institute of Fundamental Research, Pune 411 007, India}

\author{B.~Bhattacharyya}
\affiliation{National Centre for Radio Astrophysics, Tata
Institute of Fundamental Research, Pune 411 007, India}

\author{U.~Panda}
\affiliation{National Centre for Radio Astrophysics, Tata
Institute of Fundamental Research, Pune 411 007, India}

\author{B.~W.~Stappers}
\affiliation{Jodrell Bank Centre for Astrophysics,
School of Physics and Astronomy, The University of Manchester,
Manchester M13 9PL, UK}
\author{M.~A.~McLaughlin}
\affil{Department of Physics and Astronomy, West Virginia University, Morgantown, WV 26506, USA}
\affil{Center for Gravitational Waves and Cosmology, West Virginia University, Chestnut Ridge Research Building, Morgantown, WV 26505, USA}



\begin{abstract}

The fast Fourier transform (FFT) based periodicity search methods provide an efficient way to search for millisecond and binary pulsars but encounter significant sensitivity degradation while searching for long period and short duty cycle pulsars. An alternative to FFT-based search methods called the Fast Folding Algorithm (FFA) search, provides superior sensitivity to search for signals with long periods and short duty cycles. In the GMRT High Resolution Southern Sky (GHRSS) survey, we are using an FFA-based pipeline to search for isolated pulsars in a period range of 100 ms to 100 s. We have processed 2800 degree$^2$ of the sky coverage away from the Galactic plane and discovered 6 new pulsars. Here, we report the discovery of 4 of these pulsars with the FFA search pipeline. This includes a narrow duty cycle pulsar, J1936$-$30, which shows nulling behavior with an extreme nulling fraction of $\sim 90\%$. Two of the GHRSS discoveries from the FFA search lie in narrow duty cycle ranges beyond the limit of the existing population. The implementation of FFA search in the GHRSS survey and other pulsar surveys is expected to recover the missing population of long period and short duty cycle pulsars.   

\end{abstract}

\keywords{neutron stars -- pulsar surveys -- individual pulsar}


\section{Introduction} \label{sec:intro}
Neutron stars are compact stellar remnants. Radio pulsars are a type of rotating neutron stars that exhibit pulsed emission in time-domain observations at radio frequencies. Pulsars are excellent physical laboratories to probe physical processes under extreme conditions \citep{physics_lab}. Their immense density provides a mean to study the behavior of super-dense matter \citep{dense_pulsar}. The electromagnetic environment of the pulsar magnetosphere creates an ideal situation to study plasma physics under extreme conditions \citep{plasma_minhajur,minhajur_2,Plasma_allexander}. The propagation through the interstellar medium (ISM) introduces frequency dependent changes in the pulsar signal (i.e. dispersion, scattering, and Faraday rotation). These effects are in turn  utilized to study the ISM \citep{ISM_filed_pulsars,ISM_pulsars}.  

Apart from all this, the origin of radio emission from pulsars is still an unresolved problem. Even after five decades of extensive study, the radio emission mechanism from pulsars is still not fully understood \citep{Beskin_2018,Mitra_2017,Cerutti_2016}. Slow pulsars are the population of non-recycled pulsars that are typically used to study the radio emission features from pulsars. The studies of slow pulsars have revealed a number of interesting emission features like microstructure \citep{craft_1968}, nulling \citep{Backer_1970}, intermittency \citep{intermittency}, drifting \citep{drifting_drake}, mode-changing \citep{bartel_1982} and features related to polarization. These  features, containing valuable imprints about the origin of radio emission, play an important role in understanding emission physics.

Since pulsars usually appear as repeating signals in time series, they are searched for as periodic signals in the time domain observations. The usual way to search for periodic signals is to use the Fast Fourier Transform (FFT) based search, where the power spectra of the time series is searched for peaks, and the power in the harmonics are added to increase the detection significance \citep{Ransom_2002}. The FFT-based search provides a computationally efficient way to search for both accelerated (binary systems) and non-accelerated (isolated pulsars) periodic signals. The currently known pulsar population has mostly been discovered through FFT-based searches. However, FFT-based searches are known to be severely affected by `red-noise' in the telescope data, resulting in a reduction of sensitivity towards long period pulsars (\citet{Heerden_2016} and \citet{papaer1}, SS22 hereafter). Also, the incoherent summing of harmonic power over a limited harmonic space reduces FFT search sensitivity for small duty cycle (fraction of the pulse profile that contains signal from the pulsar) pulsars, as signals from such pulsars are spread over a large number of harmonics. It has been shown that irrespective of the number of harmonics added, full recovery of the power of the pulsar signal is not possible if the duty cycle of the pulsar is small \citep{RIPTIDE}. These limitations of the FFT search can bias the current pulsar population against long period, short duty cycle pulsars. An alternative search method, the Fast Folding Algorithm (FFA) search \citep{staelin}, based on folding of the time series, has a uniform response to all periods and duty cycles (\citet{RIPTIDE} and SS22). The FFA search is also known to have superior sensitivity for all periodic signals. A rigorous theoretical analysis of FFA and FFT search sensitivity has been done by \citet{RIPTIDE}. In an earlier paper (SS22), we tested the theoretical predictions by \citet{RIPTIDE} and compared FFA and FFT based searches in ideal white noise and real noise conditions with the time-domain data from the GMRT High Resolution Southern Sky (GHRSS) survey. We also presented an FFA based search pipeline\footnote{https://github.com/GHRSS/ffapipe} for the GHRSS survey along with a few initial discoveries. Many other pulsar surveys (e.g. SUPERB survey\footnote{https://sites.google.com/site/publicsuperb/discoveries}, PALFA survey\footnote{http://www2.naic.edu/palfa/newpulsars/index.html}, HTRU-S survey\footnote{https://sites.google.com/site/htrusouthdeep/}) have implemented FFA search pipelines to get optimal sensitivity for long period pulsars. Implementation of FFA search in pulsar surveys is expected to reduce the bias against long period and small duty cycle pulsars. 

The current pulsar population follows a lower boundary line (LBL) on duty cycle with an approximately $P^{-0.5}$ dependence (\citet{Posselt_2021} (see Fig. 4), \citet{Skrzypczak_2018}, \citet{Mitra_2017}, and \citet{Maciesiak_2012}). \cite{Maciesiak_2012} argue that the LBL corresponds to the narrowest features distinguishable in the average beam. These narrowest features are generated by the expanded spark associated plasma columns at the radio emission heights. The dependence of LBL on spin period is associated with the divergence of dipolar magnetic field lines containing the plasma column \citep{Maciesiak_2012}. However, \cite{Skrzypczak_2018} argue that the period dependence of profile component widths can not be explained by the period dependence of open dipolar field lines and some unaccounted physical effect is responsible. In ideal white noise conditions, the FFT search with 8 harmonic summing recovers less than 50\% of the injected power if the duty cycle is less than 1\% (reported in SS22). The presence of red-noise in the real telescope data makes the situation even worse by reducing the FFT sensitivity  for larger duty cycles as well (see Fig. 2 of SS22). Considering that most of the pulsar population has been discovered by the FFT-based search,  this lower boundary line is probably caused by the bias introduced by the FFT search against small duty cycles in addition to the reduced probability of narrow-beam pulsars pointing towards us. In a recent population synthesis work reported in  \citet{dirson_2022} (see Fig. 10), the expected median values of profile widths at longer periods were found to be smaller than that for the known population, which again indicates  a missing population of long period and small duty cycle pulsars.\\

Long period pulsars (e.g. 8.5s pulsar J2144$-$3933 \citep{young_1999}, 12.1 s pulsar 
J2251$-$3711 \citep{Morello_2020}, and 23.5 s pulsar J0250$+$5854 \citep{Tan_2018}) are always interesting objects to study in order to characterise their emission properties and constrain the pulsar death-line models \citep{single_park,Agar_2021,Tan_2018,Morello_2020}. However, only a handful of such ultra long period pulsars (P $>$ 10 s) are presently known. The deficiency of long period pulsars can be attributed to the limitations introduced by search methods, presence of red-noise, shorter integration time per survey pointing, and smaller duty cycles of such long period pulsars. A recent discovery of an ultra-long period pulsar J0901$-$4046 with a period of 76 s \citep{Caleb_2022}, has opened a new parameter space for periodicity searches.

In this paper, we report discovery of four pulsars with the FFA search pipeline. The plan of the paper is the following. In section 2, 
we describe the details of observations and briefly explain the FFA search methodology in section 3. We present the new discoveries and discuss the narrow duty cycle aspects of these FFA-discovered pulsars in section 4. In section 5, we describe the nulling features of a newly discovered pulsar reported in this paper. We summarize the work and discuss the scope of the FFA search to find the missing population of faint long period short duty cycle pulsars. 

\section{Observations}
The GHRSS survey is a survey for pulsars and radio transients conducted with the GMRT in Band-3 (300$-$500 MHz), with an incoherent array (IA) beam of the full array. The survey is divided into two phases, Phase-I with the legacy GMRT system with a narrow-band receiver (Roy et al. 2010) in the frequency range of  306$-$338 MHz \citep{GHRSS1,GHRSS2} and Phase-II with the upgraded GMRT (uGMRT) system with a wideband receiver (Reddy et al. 2017) in the frequency range of 300$-$500 MHz. The Phase-I of the survey had two time and frequency resolutions: 30.72 $\mu s$ time resolution with 16.27 kHz frequency resolution and 61.44 $\mu s$ time resolution with 32.54 kHz frequency resolution; either of the two was used depending on the maximum expected line-of-sight dispersion in the target sky region. The Phase-II of the survey has a time resolution of 81.92 $\mu s$ and a frequency resolution of 48.82 kHz. This survey targets sky away from the Galactic plane ($|b|>3^\circ$, see \citet{GHRSS1}, \citet{GHRSS2}, and SS22 for more information). The survey has discovered 28 pulsars and 2 RRATs till now, 22 pulsars with the FFT search and 6 pulsars with the FFA search. The results from the analysis of data covering 1500 deg$^2$ of the GHRSS sky, resulted in the discovery of two pulsars published in SS22. Here, we present the results from processed data covering the next 1300 deg$^2$ of the sky, resulting in 4 new discoveries.

\section{FFA search in the GHRSS survey}
We have implemented a \texttt{RIPTIDE} \citep{RIPTIDE} based FFA search pipeline in the GHRSS survey to achieve optimal sensitivity for long period pulsars.
Red-noise parameters in the data from Phase-I and Phase-II of the survey are very different. Since Phase-I data has less severe red-noise conditions, we have configured the pipeline to search in the period range 500 ms to 100 s for Phase-I and 100 ms to 100 s for Phase-II data (see SS22 for more details). 

The raw filterbank file is first cleaned to remove the frequency and time domain radio frequency interference (RFI). Then the cleaned filterbank file is dedispersed at trial DM values ranging between 1$-$500 pc cm$^{-3}$ using \texttt{PRESTO}'s \citep{Ransom_2002} \texttt{prepsubband} function and down-sampled to 1 ms time resolution. A running median subtraction is applied to the dedispersed time series by the FFA search package \texttt{RIPTIDE} to remove the red-noise contributions introducing baseline variations. After reducing the baseline undulation, the FFA search is performed on the time series using the \texttt{ffa\_search} routine of \texttt{RIPTIDE}. Finally, a peak detection on the 2D phasogram resulting from the search (see Figure 5 of SS22) followed by peak clustering is done to generate final candidates. The candidate plots generated by \texttt{RIPTIDE} do not have a frequency versus phase plot, which is needed to properly characterize the broadband nature of the pulsar signal, so we generate frequency versus phase plots by folding the cleaned filterbank files at the detected period and DM (dispersion measure, electron number density integrated along the line of sight) of the candidates. All the candidates are then visually investigated to discriminate between the pulsar and non-pulsar candidates.

\section{Discoveries}
\begin{figure}
   
    \includegraphics[height=7 cm, width=18 cm]{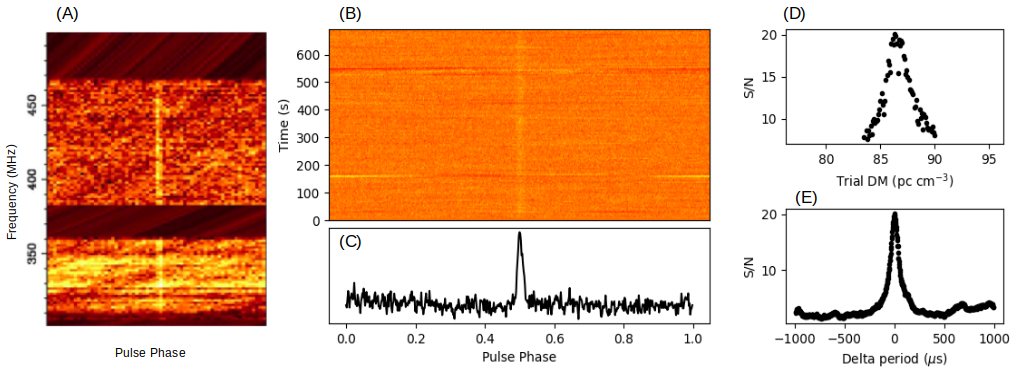}
    \caption{ Detection of PSR J1245$-$52, discovered in the GHRSS survey. This is a 835.3 ms pulsar having a DM of 86.3 pc cm$^{-3}$. (A) Frequency versus pulse phase plot, obtained from post-processing and added to RIPTIDE output for better visualization of the broadband nature of the signal. (B) Time versus pulse phase plot. (C) Integrated pulse profile. (D) Plot of achieved S/N versus trial DM, and (E) Plot of the detection S/N versus trial period.}
    \label{fig:fig1}
\end{figure}

\begin{figure}
   
    \includegraphics[height=7 cm, width=18 cm]{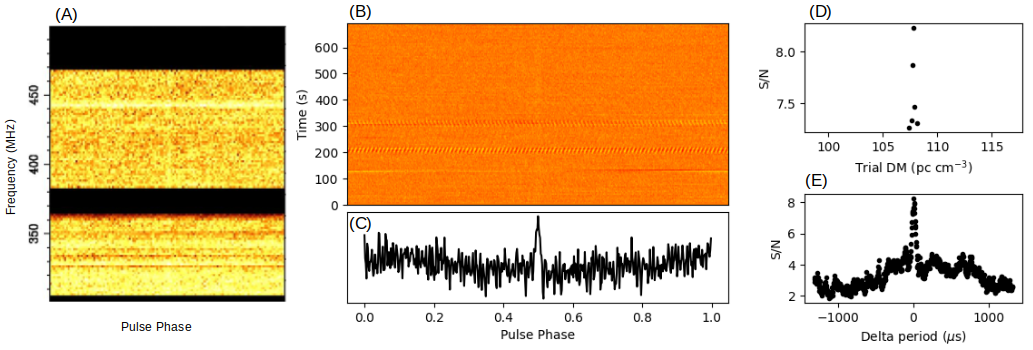}
    \caption{ Detection of PSR J1447$-$50, discovered in GHRSS survey. This is a 960.2 ms pulsar having a DM of 107.8 pc cm$^{-3}$. The panel descriptions are same as Fig. \ref{fig:fig1}.}
    \label{fig:fig2}
\end{figure}

\begin{figure}
   
    \includegraphics[height=7 cm, width=18 cm]{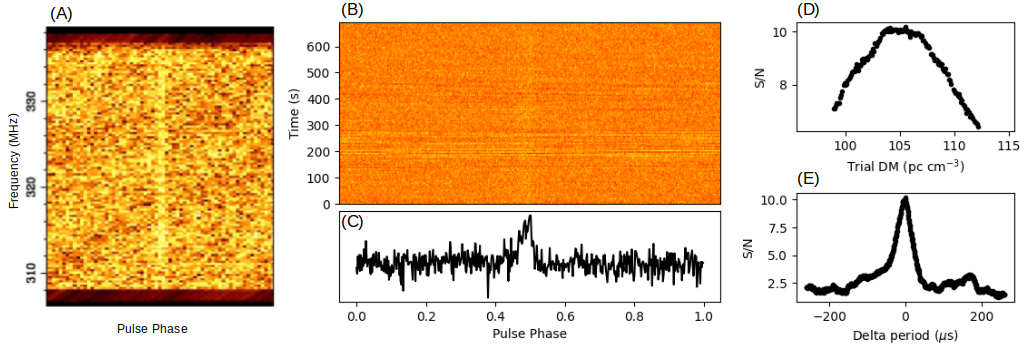}
    \caption{Detection of PSR J1810$-$42, discovered in GHRSS survey. This is a 532.2 ms pulsar having a DM of 104.9 pc cm$^{-3}$. This is the only pulsar discovered by FFA search in phase-I archival data from the survey.  The panel descriptions are same as Fig. \ref{fig:fig1}.}
    \label{fig:fig3}
\end{figure}

\begin{figure}
   
    \includegraphics[height=7 cm, width=18 cm]{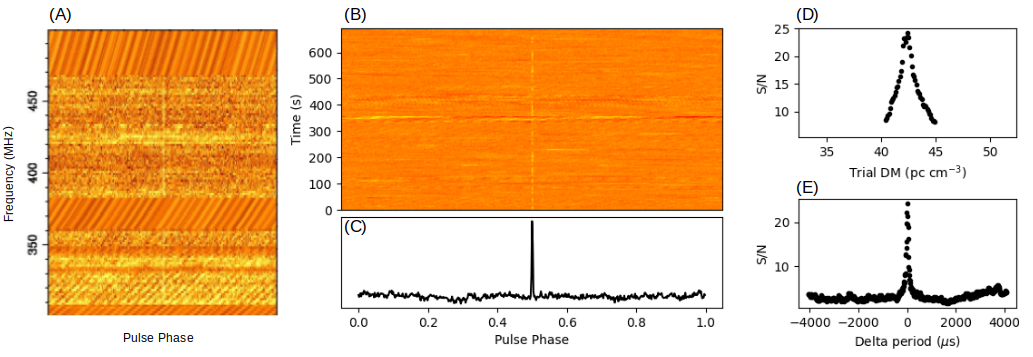}
    \caption{ Detection of PSR J1936$-$30, discovered in GHRSS survey. This is a 1675.8 ms pulsar having a DM of 42.2 pc cm$^{-3}$. This pulsar has one of the smallest duty cycle ($0.9^{\circ}$) in the known population and this also shows nulling.  The panel descriptions are same as Fig. \ref{fig:fig1}.}
    \label{fig:fig4}
\end{figure}

With the newly implemented FFA-based search pipeline, we have discovered six new pulsars till now. Two pulsars discovered from the initial processing of data covering 1500 degree$^2$ were reported in \citet{papaer1}, along with an independent discovery of J1842$-$39, a pulsar originally discovered in the GBNCC survey\footnote{http://astro.phys.wvu.edu/GBNCC/}. In this paper, we report the discovery of four pulsars from further processing of data covering 1300 degree$^2$. Table \ref{table:1} lists the parameters of all six pulsars discovered by the FFA search pipeline. We estimate the flux density of these pulsars using the radiometer equation without considering the effect of the primary beam, as the exact location of the pulsar within the primary beam is not yet known. Figures \ref{fig:fig1}, \ref{fig:fig2}, \ref{fig:fig3}, and \ref{fig:fig4} show the discovery plots of the four new pulsars reported in this paper. Fig. \ref{fig:fig3} shows the plot for pulsar J1810$-$42, which is the only pulsar discovered from reprocessing of Phase-I (narrow band) archival data. PSRs J1245$-$52 and J1447$-$50 were discovered in reprocessing of Phase-II (wide band) archival data. PSR J1936$-$30 was discovered in new data coming from the ongoing survey observations. The FFA search also helped us to get the true periods of some of the FFT discovered GHRSS pulsars that were not detected at their true periods. A GHRSS pulsar J0941$-$43 was detected at a period of 1.5664 s in the FFA search, while its earlier known period from the FFT search was 447.7 ms, which is a 7/2 harmonic of the true period.
\begin{deluxetable*}{ccccccc}
\tablenum{1}
\tablecaption{Pulsar parameters of all the six pulsars discovered in FFA search. Two pulsars in the upper part of the table were reported in SS22. The lower part of the table lists the four newly discovered pulsars reported in this work. The flux values have been estimated using the radiometer equation without considering the primary beam effect. The pointing centre represents the center of the IA beam with a half power beam width (HPBW) of $\pm 38'$ at 400 MHz.}
\tablewidth{0pt}
\tablehead{
\colhead{Pulsar name}& \colhead{DM} & \colhead{Period } & \colhead{Duty$-$cycle} &  \colhead{FFA-S/N} & \colhead{Flux} & \colhead{Pointing Centre (IA beam)}\\
\colhead{} & \colhead{(pc cm$^{-3}$)} & \colhead{(ms)} & \colhead{W50($^{\circ}$)} & \colhead{} & \colhead{(mJy)} & \colhead{(RA(hh:mm:ss), Dec(dd:mm:ss))} 
}
\startdata
J1517-31a & 51.0 & 140.6 & 5.41 & 26 & 0.6 & (15:17:51, -31:20:40) \\
J1517-31b & 61.7 & 1103.7 & 1.52 & 16 & 0.4 & (15:17:51, -31:20:40) \\
\hline
J1245-52 & 86.3 & 835.3 & 6.67 & 30 & 0.7 & (12:45:06, -52:40:00)  \\
J1447-50 & 107.8 & 960.2 & 7.1 & 10 & 0.4 & (14:47:48, -50:00:00) \\
J1810-42 & 104.9 & 532.2 & 15.1 & 17 & 1.5 & (18:10:51, -42:00:00)\\
J1936-30 & 42.2 & 1675.8 & 0.9 & 50 & 0.7 & (19:36:15, -30:00:00)\\
\enddata
\end{deluxetable*}
\label{table:1}

\begin{figure}
   
    \includegraphics[height=7 cm, width=18 cm]{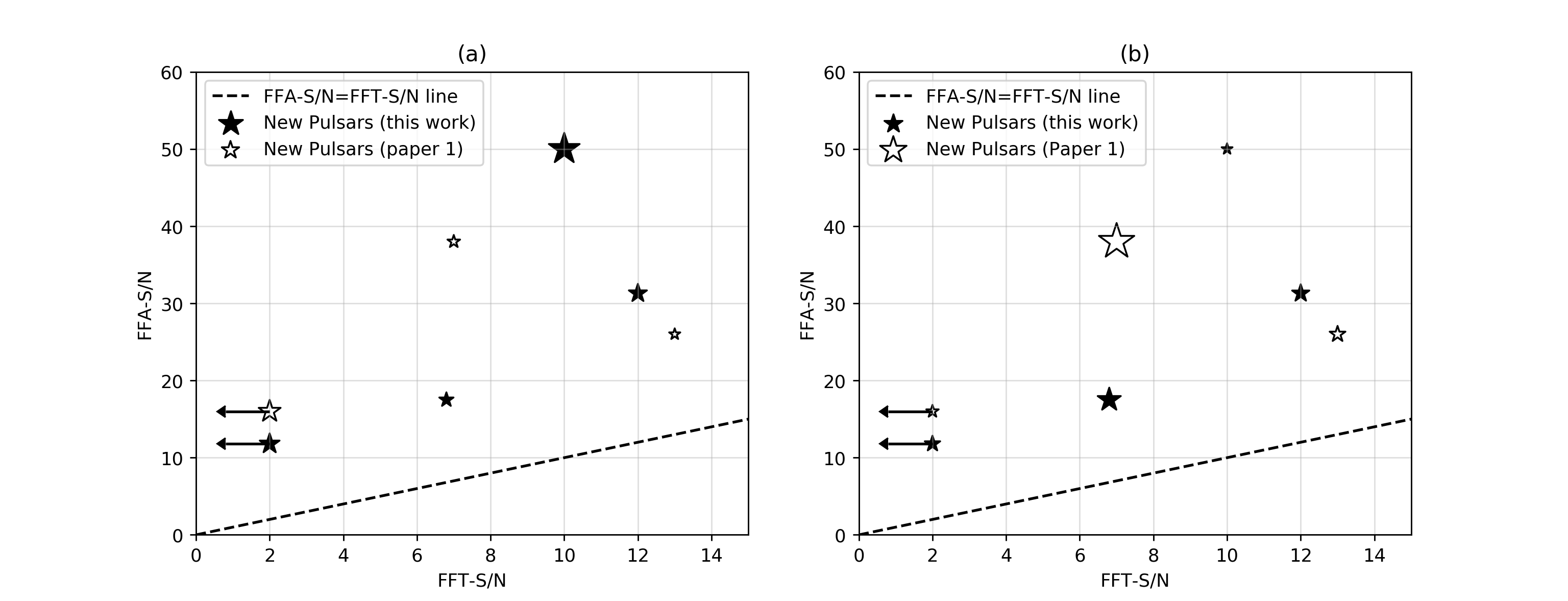}
    \caption{Comparison of FFA and FFT S/N as a function of period and duty cycle for the pulsars discovered by the FFA search in GHRSS survey. In plot (a), the size of markers is proportional to the period of the pulsar (ranging between 0.14 to 1.67 s) and in plot (b) it is proportional to the duty cycle (ranging between 0.5$\%$ to 10$\%$). The black dashed line represents a 1:1 line (where FFA S/N  and FFT S/N are equal). The empty stars are the pulsars that were reported in \citet{papaer1} (including the independent discovery of the GBNCC PSR J1842$-$39) and filled stars are the pulsars reported in this paper. Two of the new pulsars were missed by the FFT search with 8 harmonic summing at a 2$\sigma$ significance (default cut-off of \texttt{accelsearch} of \texttt{PRESTO}). These plots agree with our finding in \citet{papaer1} that all the pulsars are better detected in the FFA search.}
    \label{fig:fig5}
\end{figure}

In the latest theoretical study of FFA and FFT search sensitivity, \citet{RIPTIDE} found that FFT search sensitivity is only a function of the duty cycle and has no dependence on the period of the signal, and FFA search sensitivity does not depend on either period or duty cycle of the periodic signal. These findings are true only if ideal white noise is considered. The FFT search sensitivity is also heavily dependent on the period of the pulsar in real telescope data, due to the presence of red-noise (SS22). The dependence of FFT search sensitivity on the duty cycle of the signal also appears differently in real data than in ideal white noise conditions. While FFT search sensitivity is expected to increase with increasing duty cycle, we found that FFT search sensitivity does not improve with increasing duty cycle for GHRSS noise conditions
(see figure 2 of SS22). FFA search sensitivity with appropriate removal of the baseline variations in the time series remains uniform for all periods and duty cycles. Fig. \ref{fig:fig5} shows the FFA S/N versus FFT S/N plot for all six pulsars discovered in the FFA search along with the independently discovered PSR J1842$-$39. Pulsars that have been published in SS22 are marked by empty stars, while new pulsars reported in this paper are marked by filled stars. Marker sizes have been scaled proportional to the rotation period in panel (a) and proportional to the duty cycle in panel (b). We find that all the pulsars are better detected in the FFA search. The trend of detecting long period pulsars with more S/N in the FFA search is also visible in panel (a). Two out of six pulsars were missed by the FFT search (marked by upper limits on FFT S/N). The long period pulsar J1936$-$30, with a very narrow duty cycle, was detected at 10th harmonic with 10$\sigma$ significance in the FFT search, while it was detected at its true period with more than 50$\sigma$ significance in the FFA search. It should be noted that all the newly discovered pulsars have FFT S/N less than 14$\sigma$, while FFA S/N ranges between 10--50$\sigma$.\\

\begin{figure}
    \centering
    \includegraphics[height=10 cm, width=12 cm]{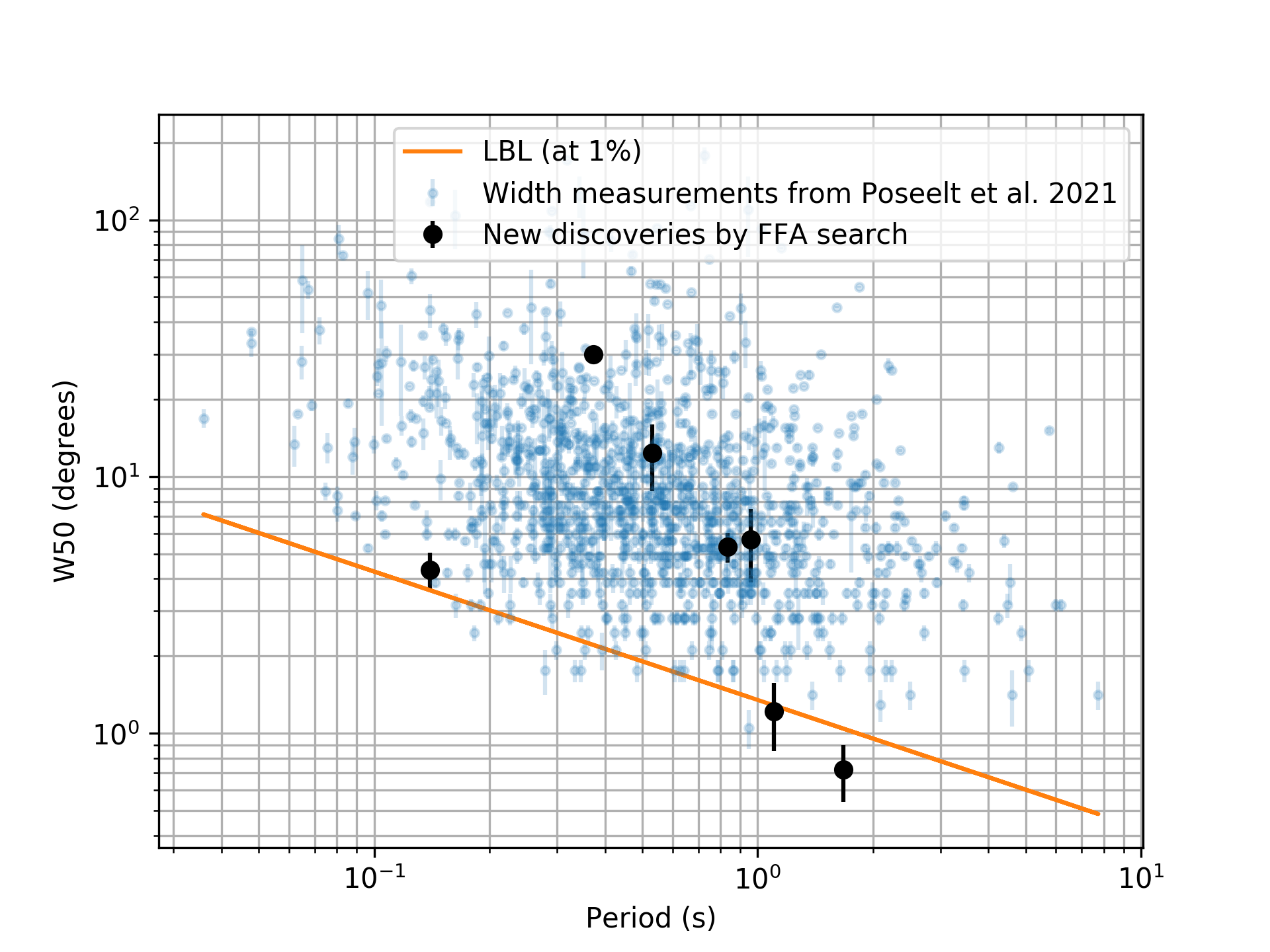}
    \caption{Duty cycle vs. period plot for the FFA discovered pulsars. The W50 values have been scaled by $\nu^{-0.19}$ (accounting for radius-to-frequency mapping) to get width at 1250 MHz from our 400 MHz values and are plotted with width measurements of 1181 pulsars from \citet{Posselt_2021}. A lower boundary line, where only 1\% of the population has widths below this line, has also been plotted. Three out of seven FFA discovered pulsars (including the independent discovery of GBNCC PSR J1842$-$39, which has the largest duty cycle in this sample) have width either on the lower boundary line or beyond the line. The error bars on black points are equivalent to the length of one profile bin.}
    \label{fig:fig6}
\end{figure}

Three out of six pulsars discovered with the FFA search in the GHRSS survey have very narrow duty cycles. Fig. \ref{fig:fig6} shows the scaled version ($\nu^{-0.19}$ scaling was used for radius-to-frequency mapping \citep{Mitra_2017}) of W50 measurements of these six pulsars and independently discovered pulsar J1842$-$39 (black markers) along with width measurements of 1181 pulsars at 1284 MHz reported in \citet{Posselt_2021} under the Thousand-Pulsar-Array (TPA) program on MeerKAT. The orange line is the lower boundary line with only $1\%$ of the population below this line. Two of the six FFA discovered pulsars (J1517$-$31b and J1936$-$30) are below this lower boundary line. PSR J1517$-$31b was missed by the FFT search, even with 32 harmonic summing (SS22). The second pulsar J1936$-$30, despite being bright, was detected at the 10th harmonic with 5-times lower significance in the FFT search compared to the FFA search. PSR~J1517$-$31a is on the lower boundary line. This pulsar (J1517$-$31a) has a period of 0.140 s and was detected in both the FFA and FFT search with good S/Ns (SS22). In summary, three out of six FFA discovered pulsars have widths that place them among the top 1\% smallest duty cycle pulsars of the total population with J1936$-$30 being an outlier. This demonstrates the capability of the FFA search to recover small duty cycle pulsars that could have missed by the FFT search.

\section{Nulling analysis of PSR J1936$-$30}
One of the new pulsars reported in this work, PSR J1936$-$30, exhibits regular nulling in the 10-minute duration discovery epoch. In follow-up observations at multiple epochs, this pulsar shows burst phases (duration in which emission from the pulsar is seen) and null phases (duration when the pulsar is nulling and no emission is seen) ranging between a few minutes to a few tens of minutes. The discovery observation was in a burst phase of the pulsar, enabling us to get good detection S/N. The pulsar is bright enough for us to see single pulses in the incoherent array mode of observation.\\
We used the method described by \citet{nulling_method} to calculate the nulling fraction of this pulsar. A running median filter is subtracted from the dedispersed time series to reduce the baseline variations. The histograms for off-pulse energy distribution and on-pulse energy distribution are generated using this zero median time series. The off-pulse energy histogram is fitted with a Gaussian and the amplitude ($A_0$) and standard deviation ($\sigma$) of the Gaussian are obtained. A scaled version of the Gaussian function, having the same standard deviation as the off-pulse energy distribution, is used to fit the negative portion of the on-pulse energy histogram and the amplitude of this scaled Gaussian ($A_1$) is noted. This scaled Gaussian represents the number of pulses that have similar on-pulse and off-pulse energy. The nulling fraction ($f_n$) is defined as the ratio of the two amplitudes ($A_1/A_0$). The errors associated with fitting   translate to an error in the nulling fraction and are calculated using the following equation from \citet{nulling_method},
\begin{equation}
    \sigma_{f_n}=\sqrt{\left(\frac{\delta A1}{A_0}\right)^2+\left(\frac{A1\delta A_0}{A_0^2}\right)^2}\\
\end{equation}
Here $\delta A_0$ and $\delta A_1$ are errors in the estimates of $A_0$ and $A_1$. We find a nulling fraction of $58\pm4.6\%$ for the discovery epoch.

\begin{figure}
   
    \includegraphics[height=7 cm, width=18 cm]{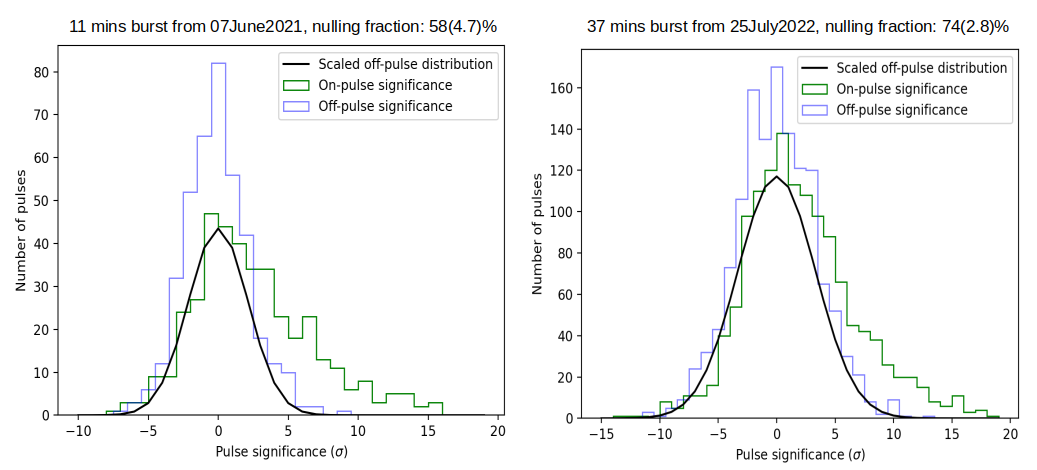}
    \caption{Energy histogram of burst phases. The left panel shows the histogram for 11.4 minutes duration of the discovery epoch (07 June 2021, MJD: 59371), while the right panel is the histogram for the longest observed burst from this pulsar on 25 July 2022 (MJD: 59785). We use the method described by \citet{nulling_method} to calculate nulling fraction in the burst phases of the pulsar. The black curve is the scaled version of the off-pulse energy distribution to fit the negative part of the on-pulse energy histogram. We find a nulling fraction of $58\pm4.7\%$ in the discovery epoch on June 2021 in 11 minutes burst and a nulling fraction of $74\pm2.8\%$ in the 37 minutes burst from the 25 July 2022 observation.}
    \label{fig:fig7}
\end{figure}

In the follow-up observations of longer duration, the pulsar shows long nulls, and the application of the nulling analysis method over the whole observation duration resulted in almost similar distributions for the on and off-pulse energies (due to the very small number of pulses in the on state). Hence, we decided to use the nulling analysis only on the burst phase of the pulsar and to include the pulse cycles of the null phase separately to calculate the overall nulling fraction for the full observation length. We also folded the null phase to confirm that there is no faint emission in the null phase. We visually identified the burst phases of the pulsar from a sub-integration plot. For sub-integration profiles of 30 s duration, the signal from the pulsar had more than 10$\sigma$ significance and were clearly distinguishable from the sub-integration with no pulsed emission. We consider an error of one sub-integration in the visual estimate of burst length. We used the nulling analysis method described in \citet{nulling_method} to calculate the nulling fraction of each burst (see Fig. \ref{fig:fig7} for single pulse energy histograms of burst phases). These nulling fractions of individual bursts are listed in Table \ref{table:2}. We use the nulling fraction of all the bursts and lengths of null phases in a given observing epoch to compute the total number of pulses with and without signal and uncertainty in this estimation. We use the ratio of the number of on-pulses and the total number of pulses in the observation as the overall nulling fraction of that observation. The error in the estimate of the number of on-pulses translates to an error in the overall nulling fraction. The overall nulling fraction of a given epoch along with its uncertainty is also listed in Table \ref{table:2}.

\begin{figure}
   
    \includegraphics[height=10 cm, width=18 cm]{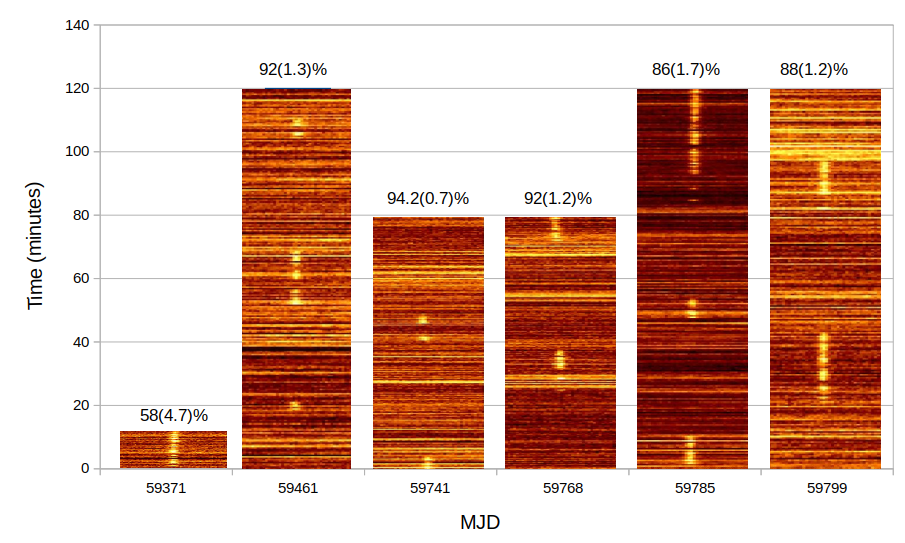}
    \caption{Pulse sequences from six epochs of observation. Each bar represents an observation, the y-axis is time in minutes, and the numbers above the bars are nulling fractions. These plots are zoomed into a phase window of $4\%$ ($15^\circ$) around the on-pulse phase. The pulsar shows bursts of emission ranging between a few minutes to a few tens of minutes, while it nulls in the remaining time. It also shows finer bursts and nulls within the long bursts along with single pulse nulls. This pulsar was discovered in a 10 minute duration of burst phase (first column in the figure), a nulling fraction of 58$\%$ was found even in this burst phase.}
    \label{fig:fig8}
\end{figure}

The nulling fraction within a burst ranges from 44$\%$ to 76$\%$, while the overall nulling fraction in sufficiently long observations (long enough to include a few null and burst phases of the pulsar, usually a few hours) is close to $90\%$, and ranges between 85$\%$-95$\%$. Fig. \ref{fig:fig8} shows the sub-integration plots zoomed in to the on-pulse phase for six epochs of observations used in Table \ref{table:2}. We note that only five known pulsars show nulling fraction larger than $85\%$ \citep{Konar_2019}. 

\begin{deluxetable*}{cccccc}
\tablenum{2}
\tablecaption{Nulling parameters of J1936$-$30 at different observation epochs. See Fig. \ref{fig:fig8} for visualization of burst and null sequences at these observation epochs.}
\tablewidth{0pt}
\tablehead{
\colhead{Epoch}& \colhead{Duration} & \colhead{Number of bursts} & \colhead{Length of bursts} &  \colhead{Nulling fraction of bursts} & \colhead{Overall nulling fraction}\\
\colhead{} & \colhead{(minutes)} & \colhead{} & \colhead{(minutes)} & \colhead{($\%$)} & \colhead{($\%$)} 
}
\startdata
59371 & 11.4 & 1 & 11.4 & 58(4.7) & 58(4.7)\\
59461 & 120 & 3 & 6, 21, 4 & 55(6), 74(3.8), 76(6.4) & 92(1.3)\\ 
59741 & 80 & 2 & 4.7, 8.4 & 44(4.1), 76(5.1) & 94.2(0.7) \\
59768 & 80 & 2 & 14, 8 & 79(3.2), 63(7) & 92(1.2)\\
59785 & 120 & 3 & 12, 7, 37 & 63(4.8), 70(6.7), 74.5(2.8) & 86(1.7)\\
59799 & 120 & 2 & 21, 10.8 & 57(3.8), 62(5.1) & 88(1.2)\\
\enddata
\end{deluxetable*}
\label{table:2}
\section{Summary}
The FFA search is now established as a more sensitive search method to find non-accelerated periodic signals in both ideal white noise as well as real telescope data
(SS22, \citet{RIPTIDE}, and  \citet{Kondratiev_2009}). The conventional FFT-based search method has a bias against long period and small duty cycle pulsars (SS22 and \citet{RIPTIDE}). Since most of the known pulsar population has been discovered by the conventional FFT search, there is a possibility of missing the population of long period and small duty cycle pulsars. This missing population can be recovered by implementing the FFA search in the major pulsar surveys.

We are processing the GHRSS survey data with the newly implemented FFA search pipeline. In this work, we report the discoveries of four new pulsars, bringing the total number of FFA discoveries in the GHRSS survey to six. We find that all the new discoveries from the FFA search pipeline have 2-6 times higher FFA detection S/N than the corresponding FFT detection S/N. Two of the six new discoveries were completely missed by the FFT search even at 2$\sigma$ significance. Most of the FFA discovered pulsars have flux densities less than 1 mJy, while most of the FFT discovered pulsars in the GHRSS survey have flux densities greater than 1 mJy. 

We find that three out of six FFA discovered pulsars have very small duty cycles and two of them are located below the lower boundary line in the duty cycle versus period plot. This highlights the capability of the FFA search algorithm to recover the population of small duty cycle, long period pulsars. One of the new pulsars reported in this paper, PSR J1936$-$30, with a period of 1.675 s, shows nulling with an extreme nulling fraction of $\sim 90\%$. Only five other known pulsars show such extreme nulling fractions \citep{Konar_2019}. An effort to localize all the new discoveries within the incoherent array (IA) beam is underway and timing solutions along with precise localization will be published soon. 

A new parameter space for search is now open after the discoveries of many interesting long period pulsars like J2251$-$3711 (12 s pulsar, \cite{Morello_2020}), J0250$+$5854 (23.5 s pulsar, \cite{Tan_2018}) and J0901$-$4046 (76 s pulsar, \cite{Caleb_2022}). Many of the long period pulsars are discovered in single pulse search (e.g. 12 s pulsar \citep{Morello_2020} and 76 s pulsar \citep{Caleb_2022}). There can be many other such long period systems that are not bright enough to be detected in a single pulse search. A periodicity search will be needed to discover such faint long period pulsars. The FFA-based periodicity search is most suited to search for such long period and small duty cycle pulsars. The discovery of 23.5 s pulsar in a periodicity search was only possible due to the very long integration time of LOFAR Tied-Array All-Sky Survey (LOTAAS)\citep{Tan_2018}. Hence, along with the implementation of the FFA search, a significant increase in the integration time per pointing is also needed to discover the population of faint long period pulsars.

\section{Acknowledgement}
We acknowledge the support of the Department of Atomic Energy, Government of India,  under project no.12-R\&D-TFR-5.02-0700.  The GMRT is run by the National Centre for Radio Astrophysics of the Tata Institute of Fundamental Research, India. We acknowledge the support of GMRT telescope operators and the GMRT staff for supporting the GHRSS survey observations. We also thank Dr. Bettina Posselt of the University of Oxford for providing the profile width measurements from the MeerKAT  Thousand-Pulsar-Array (TPA) program used in Figure \ref{fig:fig6}. MAM is supported by the NANOGrav PFC (NSF Award Number 2020265) and AccelNet (NSF Award Number 2114721) awards.

\bibliography{sample631}{}
\bibliographystyle{aasjournal}



\end{document}